\documentclass[10pt]{iopart}
\jl{4}
\usepackage[dvips]{graphicx}

\newcommand{\xp}{\ensuremath{x_{I\!\!P}}}
\newcommand{\qsq}{\ensuremath{Q^{2}}}
\newcommand{\FD}{\ensuremath{F_{2}^{D(3)}}}
\newcommand{\dcs}{\ensuremath{\sigma_{q{\bar q}}}}

\def\lsim{\mathrel{\rlap{\lower4pt\hbox{\hskip1pt$\sim$}}
    \raise1pt\hbox{$<$}}}                % less than or approx. symbol
\def\gsim{\mathrel{\rlap{\lower4pt\hbox{\hskip1pt$\sim$}}
    \raise1pt\hbox{$>$}}}                % greater than or approx. symbol

\begin{document}
\title[The colour dipole cross section]{The colour dipole approach to small $x$ processes}
\author{G R Kerley and M McDermott}
\address{Department of Physics and Astronomy,
University of Manchester, Schuster Laboratory,
Brunswick Street, Manchester, M13~9PL, England}

\begin{abstract}
We explain why it is possible to formulate a wide variety of high energy (small-$x$) photon-proton processes in terms of a universal dipole cross section and compare and contrast various parameterizations of this function that exist 
in the literature.
\end{abstract}

\section*{Introduction}

In the high-energy (small-$x$) limit of photon-proton processes there 
is a well-known (and very old \cite{gw}) factorization of time-scales 
that allows diffraction to occur. In the proton's rest frame,  
Deep Inelastic Scattering 
can be described by long lived photonic fluctuations, 
dominated by the  q\={q} Fock state, which scatter off the proton 
over much shorter time-scales than those governing the formation of the fluctuations or the subsequent formation of the hadronic final state.
When there is no net exchange of colour between the photon system 
and the proton, a large rapidity gap is observed between the two 
remnant systems 
and for small-$t$ the proton is likely to remain intact.
The kinematics are such that virtual partonic states of the photon (vector mesons, 
real photons or the continuum) may be `diffracted into existence' by the 
relatively soft interaction \cite{mp} with the colour fields of the proton 
(for recent reviews of diffraction in photon-proton collisions and extensive references see \cite{QCD_diff_rev}). 
The experimental observation of such diffractive processes constitutes one of the major achievements of the HERA experiments.

The q\={q} Fock state is a {\it colour dipole} and may be described by specifying the photon's 
light-cone wavefunction, $\psi_{\gamma} (r,\alpha)$, which depends on the 
transverse radius $r$ and the fraction, $\alpha$, of the photon's energy carried by e.g. the quark. 
The kinematics at small-$x$ are such that the dipole is frozen on the time-scale of the interaction, which may be described
by the scattering probability of a dipole in a given configuration. Because of this factorization of time-scales, it 
is reasonable to assume that the interaction itself is to a good approximation 
independent of the (much slower) formation of the dipole and of the hadronization process.
This leads to the concept of a {\it dipole cross section}, a universal quantity which may be used not only for 
all small-$x$ exclusive processes but also the inclusive photon-proton scattering via the optical theorem. 
This universal interaction is driven by the soft (i.e. low momentum fraction, or sea) structure of the proton, 
which for perturbative interactions corresponds mainly to gluons.

The suppression of large dipoles in the photon wave function at large \qsq\ means that the small $r$ and large $r$ 
regions correspond roughly to perturbative and non-perturbative contributions respectively to the overall process. 
It is not clear whether the two regimes may be separated at small $x$.
A number of authors have extracted the dipole cross section (DCS) from data and make  
predictions for other process rates~\cite{dcs_nik, vm_dcs_nik_1, vm_dcs_nik_2, sto_vac_pir, vm_dosch_1, dcs_dosch_2, dcs_dosch_1,  dcs_wust, sat_wust}. In this paper, we discuss assumptions which influence the various choices of 
parameterizations of the DCS. 

\section*{A comparison of models for the Dipole Cross Section}

The DCS is related to the inclusive photon-proton cross sections via an integral over all dipole configurations weighted 
by the square of the light-cone wavefunction of the photon of the appropriate polarization:

\begin{equation}
\sigma^{L,T}_{\gamma^{*}, p} = \int \mbox{d} \alpha \mbox{d}^{2} r |\psi_{\gamma}^ {L,T}(\alpha,r)|^{2} \dcs(s,r,\alpha)  
\end{equation} 
%where \(\psi_{\gamma}^{L,T}(\alpha,r)\) are the longitudinal and transverse components of the light cone photon wave function.

\noindent Forshaw \etal \cite{extr_dcs} have fitted a parameterization of the DCS 
\begin{equation}
  \dcs(s,r)  =  a\frac{P^{2}_{s}(r)}{1 + P^{2}_{s}(r)}(r^{2}s)^{\lambda_{s}} + b P^{2}_{h}(r)
  \exp(-\nu_{h}^{2}r) (r^{2}s)^{\lambda_{h}}  
\label{eq:fks}
\end{equation}
where $P_{s}(r)$ and $P_{h}(r) $ are polynomials in r.  The first term representing the `soft' contribution with energy exponent $\lambda_{s} =  0.06$ has a polynomial fraction that saturates at large $r$; the second, `hard' term with energy exponent $\lambda_{h} =  0.39$ provides the steep, small $x$ rise at large $Q^{2}$  and dies away at large $r$. Both have an $r^{2}$ dependence at small $r$.
For the photon wave function itself, the tree level QED expression~\cite{dcs_nik,wf} is used, modified by a Gaussian peak factor to represent non-perturbative effects in line with those observed in a generalized vector dominance analysis~\cite{x_fluct}.

Golec-Biernat and Wusthoff \cite{dcs_wust} suggested the following simple $x$-dependent saturating form for the dipole cross section:

\begin{equation}
\dcs(x,r) = \sigma_0 ( 1 - exp[-r^2 Q_0^2/ 4 (x/x_0)^\lambda ] )  
\label{eq:gbw}
\end{equation}

\noindent which has the sensible feature that it is proportional to $r^2$ at small $r$. It is constant at large $r$. A fit to the 
HERA data on DIS with $x < 0.01$, excluding charm and assuming $Q_0 = 1.0$ GeV, produced the following values 
$\sigma_0 = 23 $~mb, $x_0 = 3.0 \times 10^{-4}$, $\lambda = 0.29$ 
and a reasonable $\chi^2$. 
It seems intuitively clear that the DCS cannot depend on scaling variable $x = Q^2/W^2$ for very large dipoles 
(of the order of the pion radius and above). Either a flat energy ($W^2$) dependence, as above, or one
in agreement with the slow power-like Donnachie-Landshoff growth \cite{dl1} observed in hadronic cross sections 
would seem reasonable once the `dipole' reaches the size of the hadron 
(it is very unlikely that it will be literally be a  dipole for such large $r$: 
here one should think of $r$ as the typical transverse size of the complicated non-perturbative system). 
The small $r$ behaviour is expected on geometrical grounds: the cross section grows linearly with the transverse area of the dipole.

In fact, at small $r$ one may go further. It is well known from perturbative QCD, and has been applied to hard vector meson production (see e.g. \cite{fks12,fms}), that the dipole 
cross section is related to leading log accuracy 
to the LO gluon density: 

\begin{equation}
\dcs(x,r) = \frac{\pi^2 r^2}{3} \alpha_s ({\bar Q^2}) x'g(x',{\bar Q^2})  
\label{eq:qcd}  
\end{equation}

\noindent where $x' \gsim x$.  A model which exploits this fact has 
recently been developed \cite{fgms}. It was found that 
the large magnitude and steep small-$x$ rise of the LO gluon 
densities extracted from conventional global fits imply that unitarity will 
be violated by this QCD-improved form at small-$x$ 
because the magnitude of the dipole cross section becomes as large as the 
typical meson-proton cross section ($ \sigma_{\pi p} \approx 25 $~mb). 
Depending on the precise form of the ansatz used this will happen either 
within, or just beyond, the HERA region. 
This fact calls into question the use of the usual DGLAP leading-twist analysis for the analysis of small-$x$ structure functions 
for moderate photon virtualities (in the range $Q_0^2 < Q^2 < 10 $~GeV$^2$, which is usually consider a safe region in 
which to apply DGLAP).

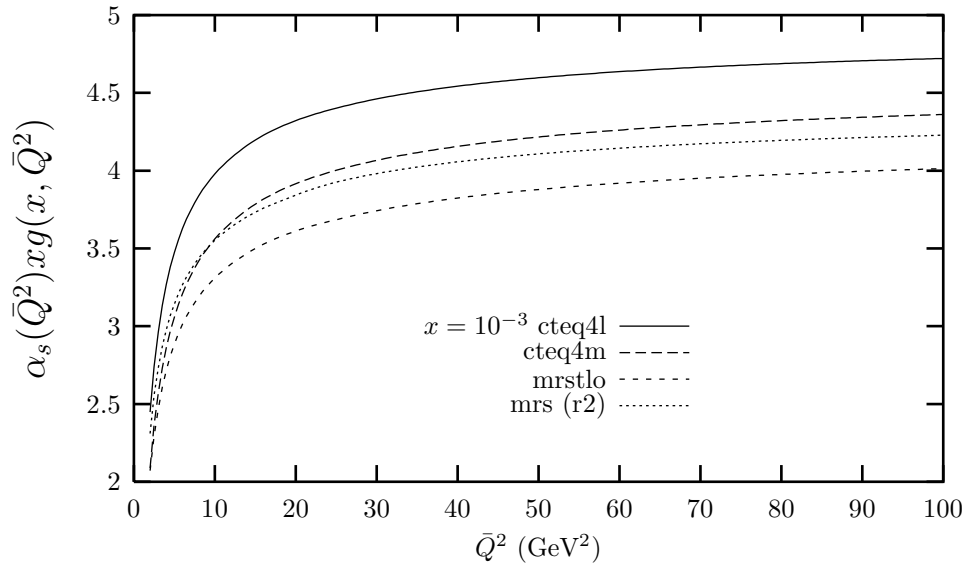
\begin{figure}[htbp]
  \begin{center}
  % GNUPLOT: LaTeX picture with Postscript
\begingroup%
  \makeatletter%
  \newcommand{\GNUPLOTspecial}{%
    \@sanitize\catcode`\%=14\relax\special}%
  \setlength{\unitlength}{0.1bp}%
\begin{picture}(3600,2160)(0,0)%
\special{psfile=asg3 llx=0 lly=0 urx=720 ury=504 rwi=7200}
\put(2180,587){\makebox(0,0)[r]{mrs (r2)}}%
\put(2180,687){\makebox(0,0)[r]{mrstlo}}%
\put(2180,787){\makebox(0,0)[r]{ cteq4m}}%
\put(2180,887){\makebox(0,0)[r]{$x = 10^{-3}$  cteq4l}}%
\put(1925,50){\makebox(0,0){${\bar Q^2}$ (GeV$^2$)}}%
\put(100,1180){%
\special{ps: gsave currentpoint currentpoint translate
270 rotate neg exch neg exch translate}%
\makebox(0,0)[b]{\shortstack{\Large  $\alpha_s ({\bar Q^2}) x g (x,{\bar Q^2})  $}}%
\special{ps: currentpoint grestore moveto}%
}%
\put(3450,200){\makebox(0,0){100}}%
\put(3145,200){\makebox(0,0){90}}%
\put(2840,200){\makebox(0,0){80}}%
\put(2535,200){\makebox(0,0){70}}%
\put(2230,200){\makebox(0,0){60}}%
\put(1925,200){\makebox(0,0){50}}%
\put(1620,200){\makebox(0,0){40}}%
\put(1315,200){\makebox(0,0){30}}%
\put(1010,200){\makebox(0,0){20}}%
\put(705,200){\makebox(0,0){10}}%
\put(400,200){\makebox(0,0){0}}%
\put(350,2060){\makebox(0,0)[r]{5}}%
\put(350,1767){\makebox(0,0)[r]{4.5}}%
\put(350,1473){\makebox(0,0)[r]{4}}%
\put(350,1180){\makebox(0,0)[r]{3.5}}%
\put(350,887){\makebox(0,0)[r]{3}}%
\put(350,593){\makebox(0,0)[r]{2.5}}%
\put(350,300){\makebox(0,0)[r]{2}}%
\end{picture}%
\endgroup
 
    \caption{The function $f = \alpha_s x g$ at fixed $x$ for various parton sets}
    \label{fig:1}
  \end{center}
\end{figure}

What should be used for ${\bar Q^2} (r^2)$ in a particular hard process? To leading-log accuracy all choices of
hard scales are equivalent, but can we make an inspired choice based on our knowledge of the $r$-integral, pending a full NLO calculation? 
 In \cite{fks12} the $r$-space expression for $F_L$  is used to set the relationship between 
transverse dipole size and four-momentum scales using the ansatz: \mbox{\(Q^2 <\!r\!>_L^2 = \lambda\)},  
where the constant $\lambda \approx 10$ is assumed to be universal. 
This clearly requires a definition of what is meant by an `average': for example one may use a median or a mode average.
For the median average, it was found in \cite{fms} that the value of $\lambda$ shows some $x,Q^2$ variation.
However, for hard enough processes the principal $r$-behaviour of the DCS comes from the $r^2$ piece in equation (\ref{eq:qcd}) 
since $\alpha_s xg$ is a rather weak function of its argument, providing the latter is large, which implies a weak $r$-dependence. 
Figure (\ref{fig:1}) illustrates the point showing this function,  at fixed $x =10^{-3}$,  for a variety of PDF sets. 
For large arguments the DCS varies little with scale, which corresponds to a weak dependence on the precise value of $\lambda$.
However it does inherit the steep energy rise from the gluon.  This observation 
would seem to support the steep small-$x$ rise implied at small $r$ in equation (\ref{eq:gbw}), 
as well as the recent two-Pomeron picture of Donnachie and Landshoff \cite{dl2} 
in which the soft Pomeron is `higher-twist' (i.e. its influence dies off with increasing $Q^2$, relative to the 
`hard Pomeron' which has a steeper energy dependence.)

The scale at which to sample $\alpha_s  xg$ under the integral is ${\bar Q^2} = \lambda/r^2$ and the 
typical average scale changes from process to process depending on the weighting 
provided by the appropriate light-cone wavefunctions under the integral in $r$. Once the relationship between transverse sizes and 
$Q^2$-scales has been set it is necessary to decide how to extrapolate the QCD-improved DCS into the non-perturbative region 
(see \cite{fgms} for more details). This inevitably leads to some model-dependence of predictions for particular small-x processes 
which is intimately related to the interplay of long and short distance Physics.

\section*{Hard exclusive diffractive processes}

This question of scale setting is very important in making pQCD predictions for hard exclusive diffractive processes such as
Hard Vector Meson Production and Deeply Virtual Compton Scattering (for which first data will soon be available). 
In these processes $\dcs$ appears in the amplitude and 
the issue of skewedness of the amplitude also plays a role and in general modifies 
the gluon density in equation (\ref{eq:qcd}) to the skewed gluon density and thereby restricting the universality of the DCS to some extent.
A topical review of recent HERA data on Vector Mesons may be found in the preprint by Crittenden \cite{crit}. 
The recent review of Teubner \cite{teub} presents a rather optimistic view of the ability of two-gluon model of 
perturbative QCD to make clear predictions for such hard diffractive processes. 
The predictions are clear in the asymptotically hard limit. However, it is not clear how close the measured data are to this limit.  
The analysis of the relatively light $J/\psi$-meson family will play a vital role because the typical scales involved in the gluon density lie in precisely the dangerous region of rather low ${\bar Q^2}$, 
where,  at fixed $x$,  $\alpha_s xg$ has strong sensitivity to its argument (see figure (\ref{fig:1})).   
At low scales this function starts off roughly flat in energy and steepens 
as the scale hardens.
This leads to very large uncertainties in both the normalization and energy dependence of processes which sample the gluon density at relatively low scales. 
As such, one of us (MM) strongly disagrees with the popular statement that 
diffractive $J/\psi$ photo- and electroproduction is well understood 
from the point of view of perturbative QCD.

Experimental evidence backs up this caution. For example, as recently pointed out by Hoyer and Peigne \cite{hoyer}, the pQCD predictions 
for the ratio of $\psi (2s) /\psi(1s)$ in photoproduction are too small  by a factor as much as five.  This conclusion
may be directly inferred by extending the analysis of the scaling issue in \cite{fms}, to smaller effective scales. 
In the opinion of one of the authors (MM) a complete, careful reanalysis of the $J/\psi$-family is urgently required, 
given a reasonable ansatz for the DCS.

\section*{Inclusive Diffraction}

\begin{figure}[htb]
\begin{center}
(a) 
\includegraphics[height=9cm,width=9cm]{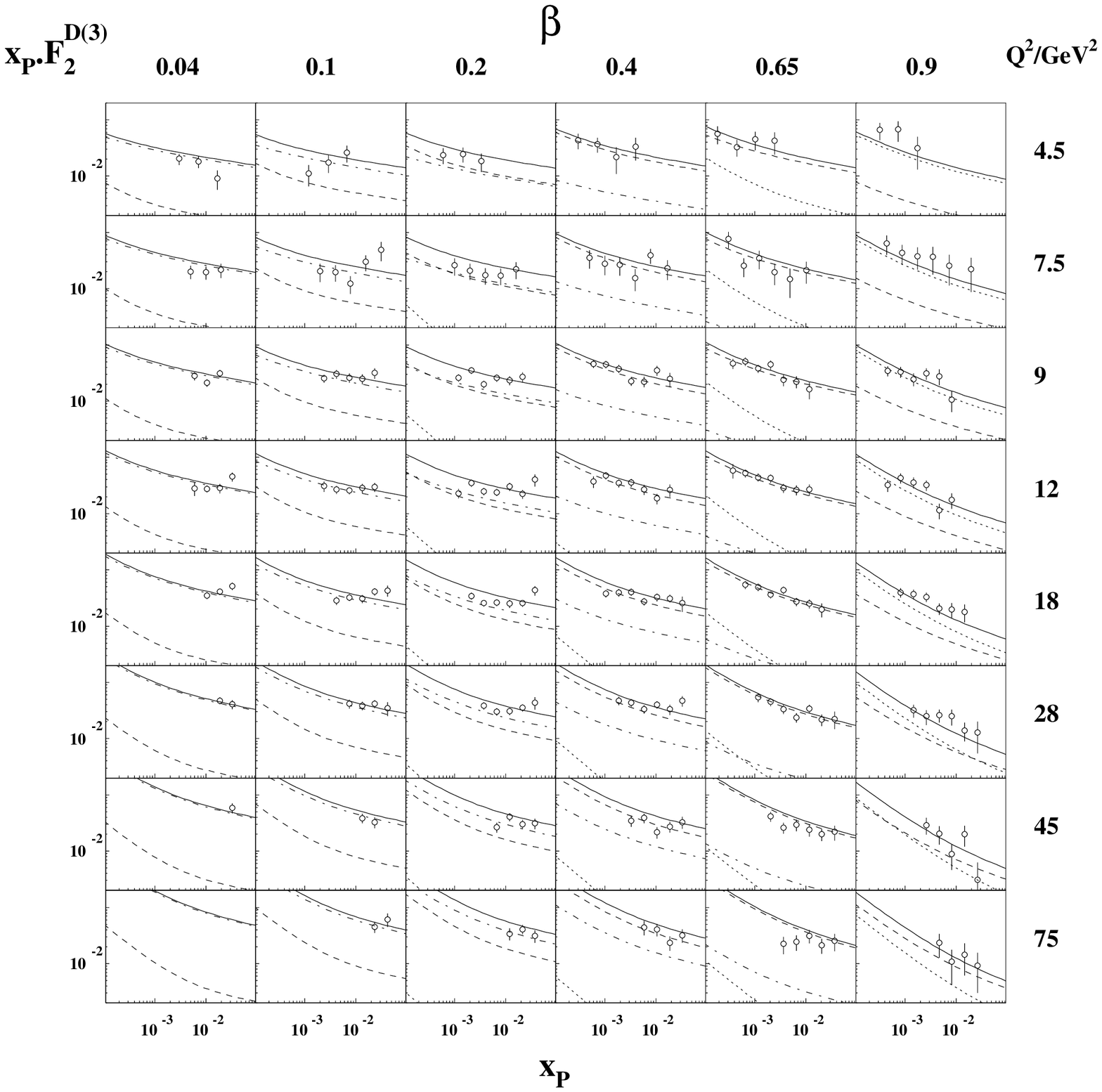}
(b) 
\includegraphics[height=9cm,width=9cm]{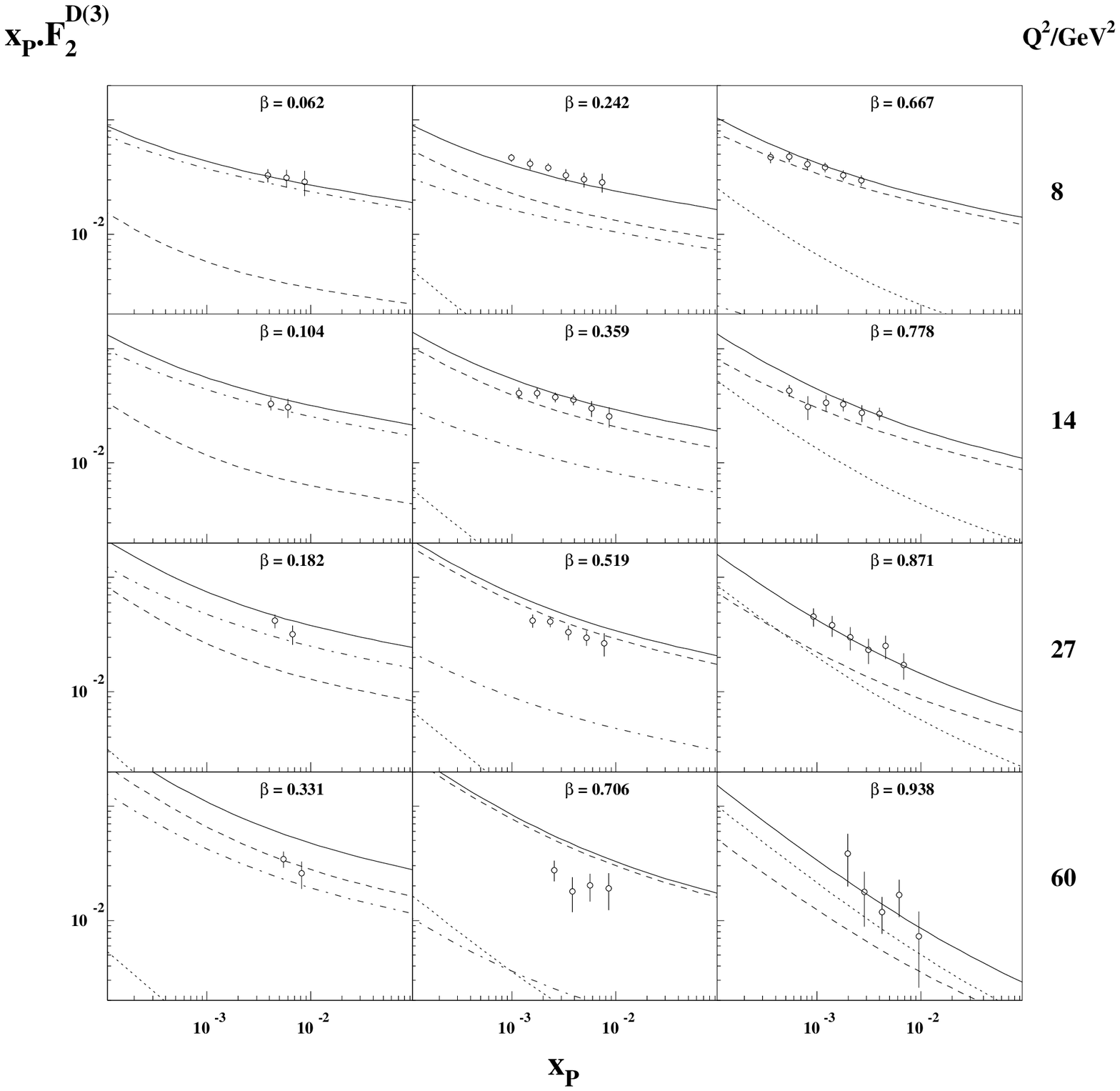} 
 \end{center}
  \caption{Contributions to $\xp \FD$ compared with (a) H1 1994 and (b) ZEUS 1994 data. 
Full, dotted, dashed and dot dashed lines are the total, longitudinal q\={q}, transverse q\={q} and 
q\={q}g contributions respectively.}
  \label{fig:f2d3}
\end{figure}

In exclusive processes, such as the diffractive structure functions, higher order Fock state are known to contribute. However 
for specific diffractive final states such as exclusive Heavy Vector Meson Production and DVCS their effects might 
be expected to be minimal.
The q\={q} dipole and q\={q}g higher Fock state contributions to $\FD$ have been calculated~\cite{dcs_f2d3} using  expressions  
derived from a momentum space treatment, with an effective two gluon dipole description for the latter~\cite{sat_wust,LC_wust}. 
Plots of these quantities are compared with H1~\cite{H1} and ZEUS~\cite{ZEUS} 1994 data in figure (\ref{fig:f2d3}).

Agreement is good for the H1 data, even at low $\beta$ where the q\={q}g term dominates.  The ZEUS data also give good agreement overall but with deviations at larger \qsq\ values for small and moderate $\beta$.

\section*{Conclusions}
In conclusion, it is still unclear whether saturation in energy is a feature of \dcs\ at present energies.
The best way to settle the issues surrounding the precise form of the DCS is to perform global analysis of all available 
small-$x$ processes: inclusive and diffractive structure functions, exclusive vector meson production, DVCS, etc. 
Different processes are sensitive to different regions in $r$, a global analysis
of this kind would therefore be very valuable in differentiating between the different ans\"{a}tze for the dipole cross section.
Obviously, the increased precision of the small-x data which should result from the forthcoming luminosity upgrade will 
be vital in addressing the Physics issues discussed here.
\ack
GRK would like to thank PPARC for a Studentship. 
MM wishes to thank Mark Strikman and Jochen Bartels for useful discussions.
\end{document}